\documentclass[pss]{wiley2sp} % provides new 2008 pss two-column style (no alternative manuscript style output available at present)
\usepackage{amsmath}
\usepackage{amssymb}
\begin{document}

% Title of the article
\title{Low temperature properties of the Electron Spin Resonance in YbRh$_2$Si$_2$}

% Abbreviated title for the page headers
\titlerunning{Low-$T$ ESR in YbRh$_2$Si$_2$}

% Authors
\author{%
  J. Sichelschmidt\textsuperscript{\Ast,\textsf{\bfseries 1}},
  T. Kambe\textsuperscript{\textsf{\bfseries 2}},
  I. Fazlishanov\textsuperscript{\textsf{\bfseries 3}},
  D. Zakharov\textsuperscript{\textsf{\bfseries 4}},
  H.-A. Krug von Nidda\textsuperscript{\textsf{\bfseries 4}},
  J. Wykhoff\textsuperscript{\textsf{\bfseries 1}},
  A. Skvortsova\textsuperscript{\textsf{\bfseries 5}},  
  S. Belov\textsuperscript{\textsf{\bfseries 5}},
  A. Kutuzov\textsuperscript{\textsf{\bfseries 5}},
  B.I. Kochelaev\textsuperscript{\textsf{\bfseries 5}},
  V. Pashchenko \textsuperscript{\textsf{\bfseries 6}},
  M. Lang \textsuperscript{\textsf{\bfseries 6}},
  C. Krellner\textsuperscript{\textsf{\bfseries 1}},
  C. Geibel\textsuperscript{\textsf{\bfseries 1}},
  F. Steglich\textsuperscript{\textsf{\bfseries 1}}
    }

% Abbreviated list of authors for the page headers
\authorrunning{J. Sichelschmidt et al.}

%E-mail-address of corresponding author
\mail{e-mail
  \textsf{Sichelschmidt@cpfs.mpg.de}, Phone:
  +49-351-46463221, Fax: +49-351-46463232}

% author's affiliations/addresses
\institute{%
  \textsuperscript{1}\,Max Planck Institute of Chemical Physics of Solids, 01187 Dresden, Germany\\
  \textsuperscript{2}\,Okayama University, Department of Physics, Okayama 700-8530, Japan\\
  \textsuperscript{3}\,E.K. Zavoisky Physical-Technical Insitute, 420049 Kazan, Russia\\
  \textsuperscript{4}\,Experimental Physics V, Center for Electronic Correlations and Magnetism, University of Augsburg, 86159 Augsburg, Germany\\
  \textsuperscript{5}\,Theoretical Physics Departement, Kazan State University, 420008 Kazan, Russia\\
  \textsuperscript{6}\, Physikalisches Institut, Goethe-Universit\"at Frankfurt, 60438 Frankfurt(M), Germany
  }

\received{XXXX, revised XXXX, accepted XXXX} % do not change, will be filled in by the publisher
\published{XXXX} % do not change, will be filled in by the publisher

%Please select four to six PACS-codes from the enclosed list (PACS.txt) or from www.aip.org/pacs)
\pacs{76.30.Kg, 71.27.+a } % For example: 71.20.Ps

\abstract{%
% This is a macro for the typesetting of two-column text in an
% abstract. It will typeset the two arguments in \abstcol{}{} as the
% left and right column inside the abstract box. At the
% columnbreak there will be always a columnbreak (\par), so both
% columns start with a new paragraph. No automatic column height
% balancing is done.
%
% If used with a \titlefigure it will silently output both
% parameters as consecutive paragraphs.
%
% The macro is defined exclusively inside the argument of \abstract{};
% if used outside it will raise an error.
%
% Usage: \abstcol{<left column>}{<right column>}
\abstcol{%
We present the field and temperature behavior of the narrow Electron Spin Resonance (ESR) response in YbRh$_{2}$Si$_{2}$ well below the single ion Kondo temperature. The ESR $g$- factor reflects a Kondo-like field and temperature evolution of the Yb$^{3+}$ magnetism. Measurements towards low temperatures ($>0.6$K) have shown}{%
distinct crossover anomalies of the ESR parameters upon
approaching the regime of a well defined heavy Fermi liquid. Comparison with the field dependence of specific heat and electrical resistivity reveal that the ESR parameters can be related to quasiparticle mass and cross section and, hence, contain inherent heavy electron properties. }}

% The class file requires the standard graphicx Latex package. See the 'LaTeX
% standard graphics and color packages documentation' for more information at
% <http://tug.ctan.org/tex-archive/macros/latex/required/graphics/grfguide.pdf>.
%
% Accepted figure file formats depend on which LaTeX flavour is used.
% Classic LaTeX is always able to use Encapsulted Postscript (EPS);
% PDFLaTeX can't use this but accepts PDF, JPG, PNG, and GIF formats.
%
% See examples for implementing graphics in floating figure environments later in this file.
% If \titlefigure is given, it takes as its mandatory parameter the
% name (without extension) of some figure file.
%\titlefigure[height=3.1cm]{empty2w}
%\titlefigurecaption{%
%  This is the caption of the \emph{optional} abstract figure. If
% there is no abstract figure here, the abstract text should be divided into both columns.}

\maketitle   % please do not remove

\section{Introduction}
The heavy fermion metal YbRh$_{2}$Si$_{2}$ has proven to display a variety of unusual low temperature electronic properties which are related to the interplay between the Kondo interaction of Yb$^{3+}$ 4$f$ spin / conduction electron spin and the indirect magnetic RKKY interaction of the 4$f$ spins. It is located very close to a magnetic instability where a weak antiferromagnetic long range order below 70~mK is suppressed by a magnetic field of $B_c=60$~mT. In the vicinity of such a magnetic field induced quantum critical point and at low temperature ($T$) pronounced non-Fermi liquid behavior is observed as evidenced by a divergence of the electronic speciÞc heat and electrical resistivity $\rho\propto T$. At higher fields electronic \mbox{specific} heat $C$, magnetic susceptibility, and $\rho$ show Landau-Fermi liquid (LFL) behavior with a renormalized electronic mass and quasiparticle scattering  \cite{gegenwart08a}. 
YbRh$_{2}$Si$_{2}$ is one of a few Kondo lattice compounds where a well-defined Electron Spin Resonance (ESR) signal allows to directly characterize the spin dynamics of the Kondo ion. Even well below a thermodynamically defined single ion Kondo temperature ($T_{\rm K}\simeq 20$~K) this signal shows properties typical for localized 4$f$ moments\cite{sichelschmidt03a,sichelschmidt07b}. It has been shown that the presence of strong ferromagnetic correlations between the 4$f$-spins enables an ESR signal narrow enough to be observed \cite{krellner08a}. Here we investigate the properties of this signal at low temperatures and as a function of low fields, below and above $B_c$.
%
%\begin{equation}
%\label{eq1}
%\frac{a}{b}=\frac{c}{d}
%\end{equation}
%
%\begin{changed}
%  This is a text snippet marked as \emph{changed}.
%  This is done by enclosing it in an environment called \verb+changed+. Please note
%  that in certain circumstances there might be small side effects such
%  as make up deviations or additional blanks.
%\end{changed}
%
\begin{figure*}[h]%
%\sidecaption
\includegraphics*[width=0.43\textwidth]{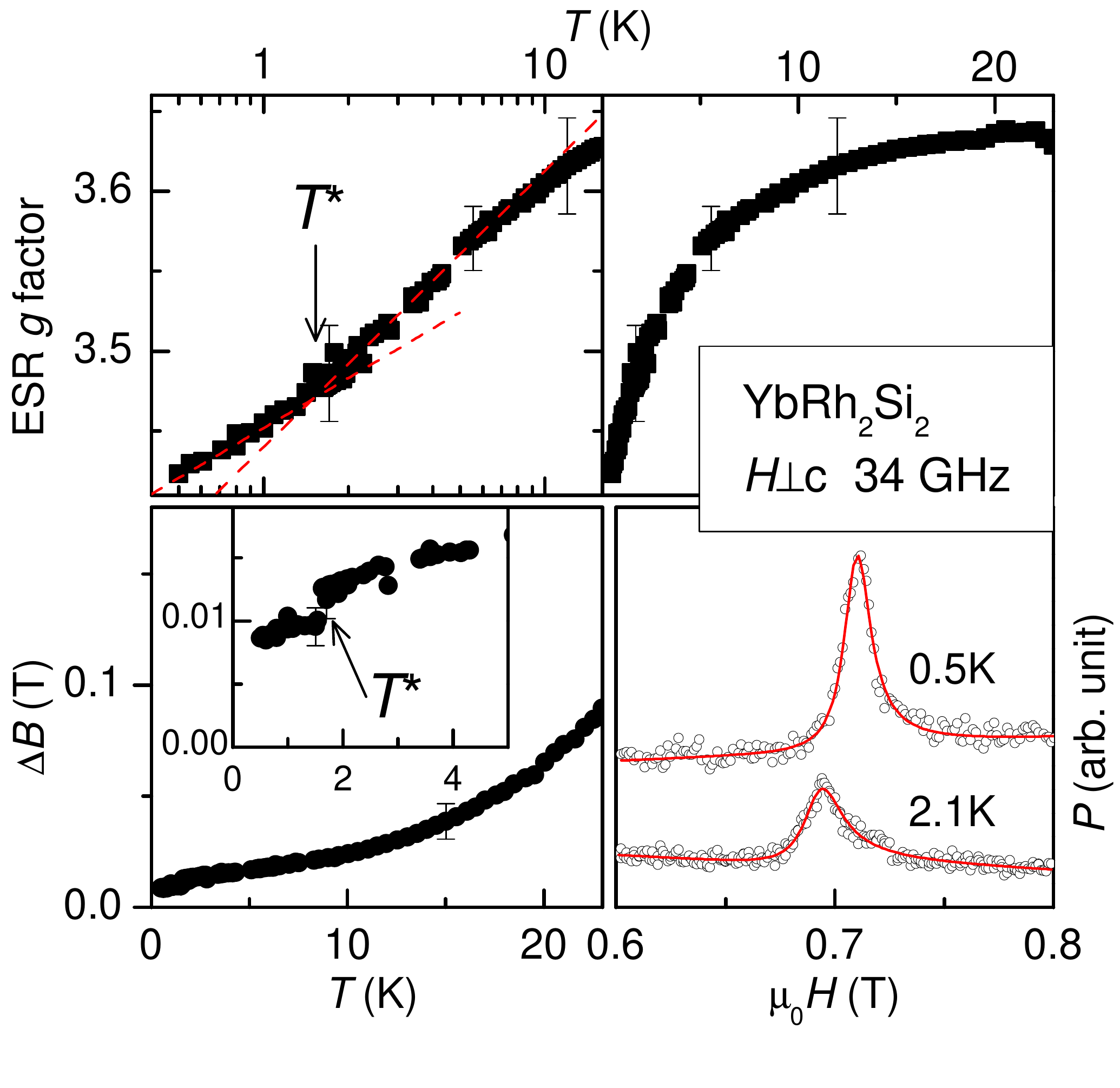}
\caption{Low temperature behavior of the Q-band (34 GHz) ESR parameters linewidth ($\Delta B$) and $g$ factor (dashed lines: $g\propto\ln T$). $T^*$ indicates the temperature of the crossover into a heavy Landau Fermi Liquid state, see the $B-T$ phasediagram in \cite{gegenwart08a}. Lower right panel shows ESR spectra ($P$: absorbed microwave power) at indicated temperatures fitted by Lorentzian lineshapes (solid lines).
}
\label{Fig1}
\end{figure*}

\section{Experiment}
We present results of ESR measurements with standard continuous wave techniques at four microwave frequencies ($\nu=1.1, 9.4, 34, 56$~ GHz) which according to the resonance condition $h\nu=g\mu_{\rm B}B_{\rm res}$  correspond to the resonance fields $B_{\rm res}=0.02, 0.19, 0.68, 1.1$~T providing the field dependence of the ESR parameters. The sample was cooled by a $^4$He flow cryostat ($T>3$~K), a $^4$He bath cryostat \mbox{($4>T>1.5$~K),} and a $^3$He bath cryostat \mbox{($2>T>0.6$~K) \cite{kajiyoshi06a}}.
As was shown previously \cite{wykhoff07a} samples with different resistivity ratios (RR~$\equiv \rho(300{\rm K})/\rho(1.8{\rm K})$) show differences in their ESR response and, therefore, we used the same  single crystal of YbRh$_{2}$Si$_{2}$ with RR=20. We ensured a proper crystal alignment (external field $H$ within tetragonal a,b-plane; $H\perp\rm c$) by minimizing the highly anisotropic resonance field. All spectra could be nicely fitted with a single Lorentzian line (containing dispersive contributions due to a finite penetration depth \cite{wykhoff07b}) as shown in the inserted frame of Fig.\ref{Fig1}.   
\section{Results}
\begin{figure*}[h]%
%\subfloat[ESR g factor. Dashed lines denote a logarithmic behavior of $g$.]{%
\includegraphics*[width=.33\textwidth]{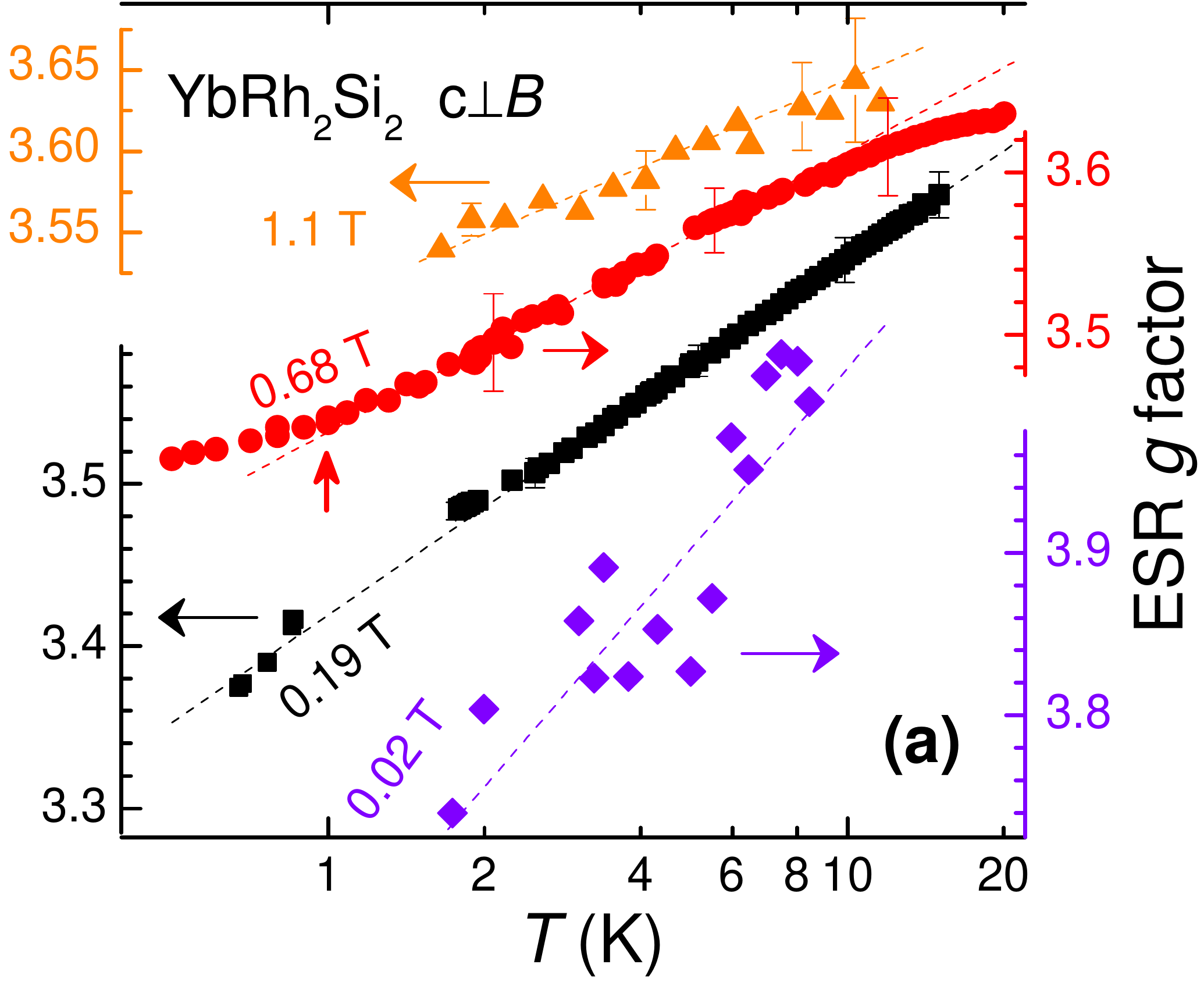}\hfill
%\subfloat[ESR linewidth $\Delta B$. Dashed lines denote a linear function emphasizing a Korringa-like relaxation.]{%
\includegraphics*[width=.33\textwidth]{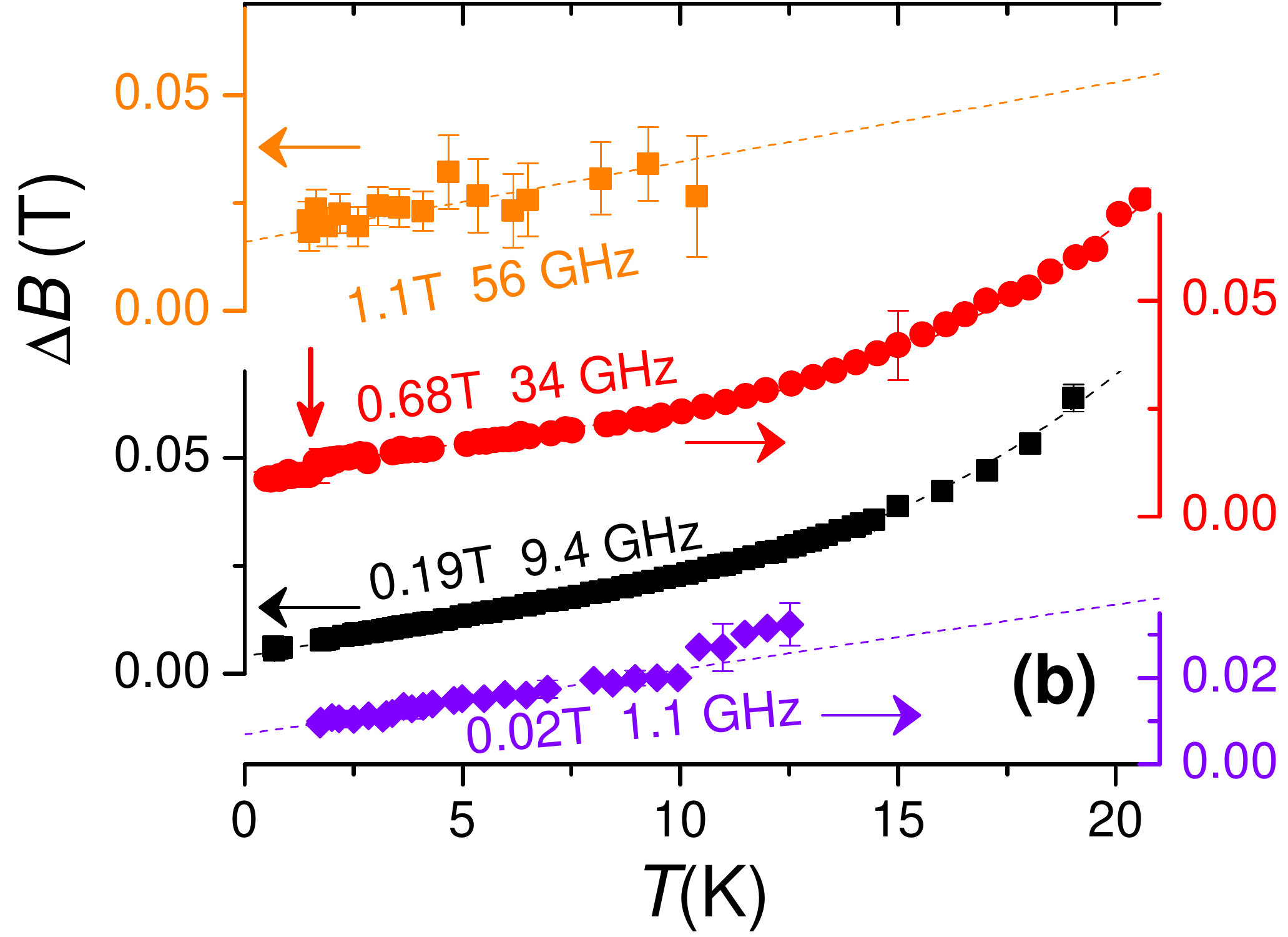}\hfill
%\subfloat[Parameters of $\Delta B=a+b\cdot T$. Dashed line indicates $a\propto B^2$.]{%
\includegraphics*[width=.3\textwidth]{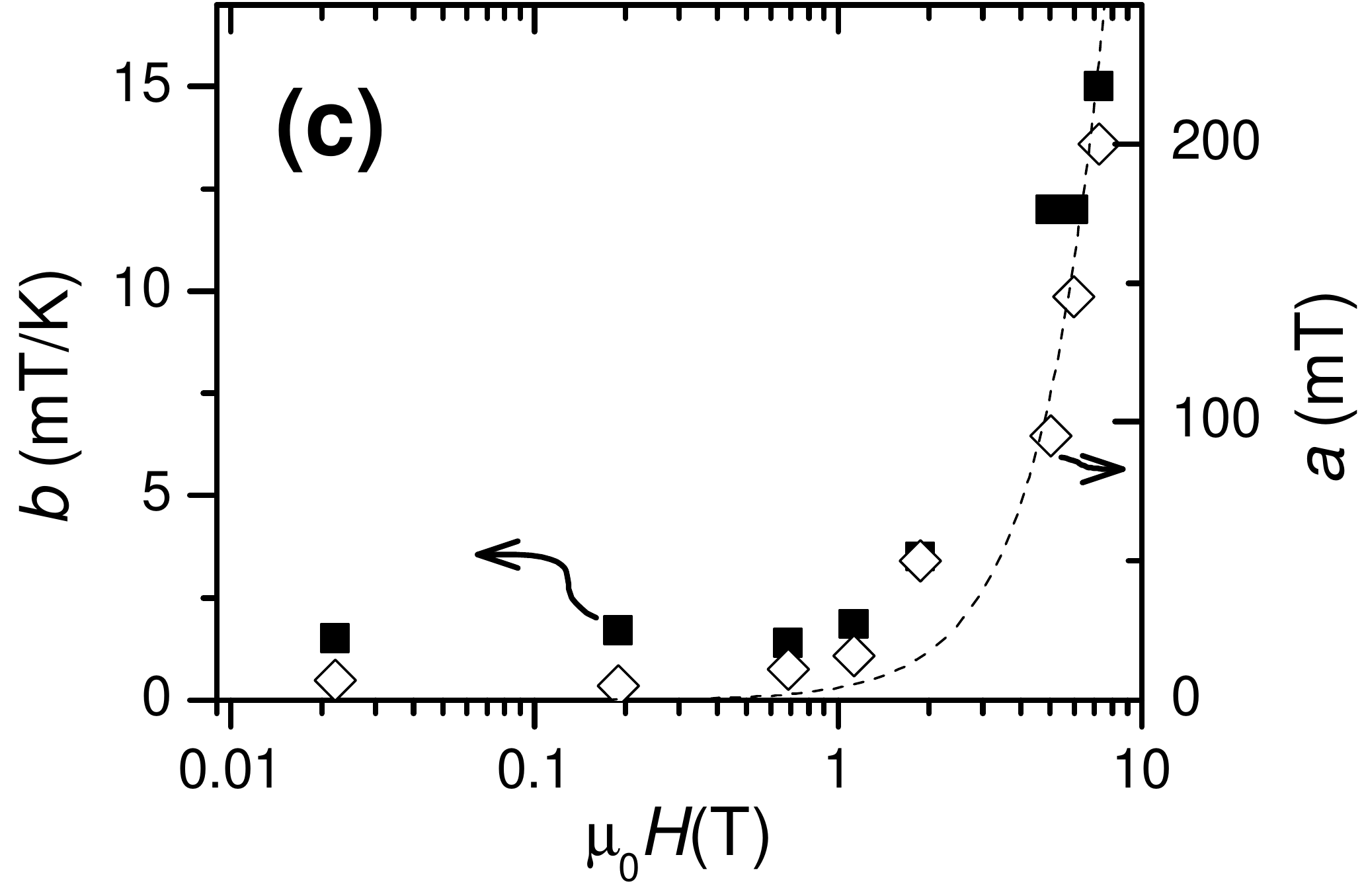}
%
%\subfloat[Third subfigure.]{\label{figs4c}%
%\includegraphics*[width=.45\textwidth,height=2.5cm]{empty2w}}%
\caption{%
Temperature dependence of (a) the ESR $g$ factor (lines: $g\propto\ln(T)$) and (b) the linewidth $\Delta B$ (dashed lines see main text). Data in (a) and (b) are depicted for various microwave frequencies corresponding to resonance fields measured at $T=5$~K. Vertical arrows indicate $T^*$, compare Fig.\ref{Fig1}. Frame (c) displays $a, b$ of $\Delta B=a+b\cdot T$ (linear part of lines in (b), data for $\mu_0H>1.8$~T from \cite{schaufuss09a}). Dashed line indicates $a\propto B^2$.
}
\label{Fig2}
\end{figure*}

The transport and magnetic properties of YbRh$_{2}$Si$_{2}$ are considerably affected below $T_{\rm K}$ by external magnetic fields up to 10~T \cite{gegenwart08a}, reflecting the strong interaction of the conduction electron spins with the Yb$^{3+}$ 4$f$ electrons which determine the magnetism with pronounced local properties \cite{kutuzov08a}. Below a characteristic temperature $T^*$ the Sommerfeld coefficient of the specific heat and the magnetic susceptibility approach Landau Fermi liquid behavior. Fig. \ref{Fig1} displays the behavior of the ESR response above and below $T^*$: Above $T^*$ the ESR $g$-factor logarithmically increases, reflecting the Kondo interaction between 4$f$ and conduction electrons \cite{sichelschmidt03a} (see below for the other Kondo-like features of $g$). As shown by the dashed lines in Fig. \ref{Fig2}b the linewidth shows a Korringa-like increase linear in temperature followed by an exponential increase due to spin-phonon interaction caused by crystal field modulations via lattice vibrations \cite{wykhoff07a}. When crossing $T^*$ from above the slope of the logarithmic $g$-factor decreases and the linewidth shows a step-like decrease. Similar, but less pronounced behavior has also been found for higher frequencies (i.e. higher magnetic fields) \cite{schaufuss09a}.\\
The temperature behavior of both $g$ factor and linewidth strongly depend on the magnetic field. Fig.\ref{Fig2} shows this dependence that we measured by using four microwave excitation frequencies corresponding to four resonance magnetic fields. The ESR $g$ factor offers a characteristic Kondo-like behavior, namely a weakening of the logarithmic temperature dependence when the magnetic field is increased, further confirming the finding of Ref. \cite{schaufuss09a} with two additional frequencies 1.1GHz and 56 GHz, respectively.
\begin{figure*}[h]%
%\sidecaption
%\begin{center}
%\subfloat[$\delta g=g(T=2\,{\rm K})-g(T=20\,\rm K)$. The data of $\delta g$ for $\mu_0H>1.8$~T are taken from Ref.\cite{schaufuss09a} ]{%
\includegraphics*[width=.35\textwidth]{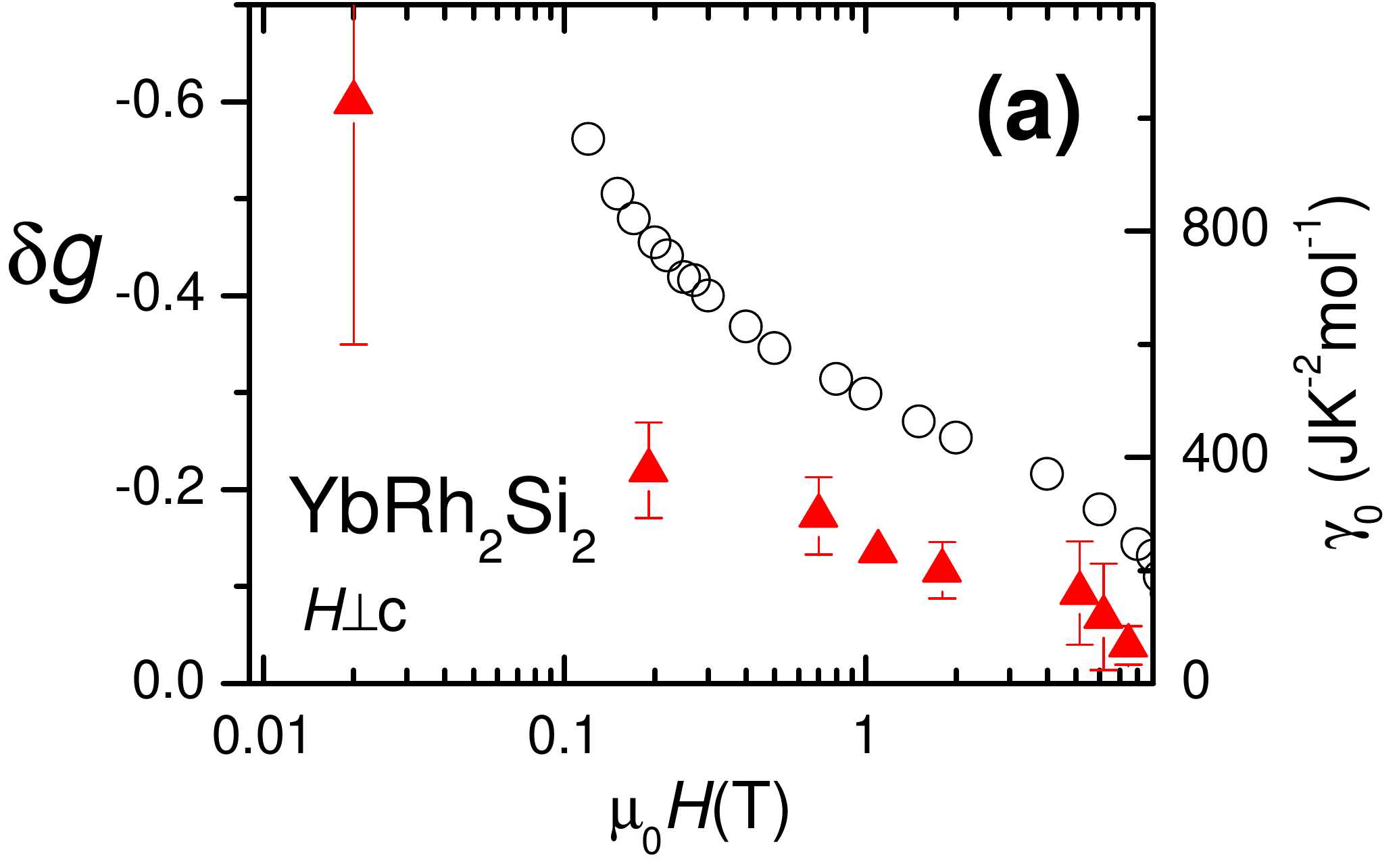}
%\hspace{6ex}
%\subfloat[Relaxation time $\tau_{\rm eff}=B_{\rm res}/(\nu\Delta B)$ obtained from the resonance field $B_{res}$ and linewidth $\Delta B$ at $T=1.8$K. The datapoint of $\tau_{\rm eff}$ for $\mu_0H=0.086$~T is obtained from a linear extrapolation of the linewidth data in Ref.\cite{duque09a}.]{%
\includegraphics*[width=.33\textwidth]{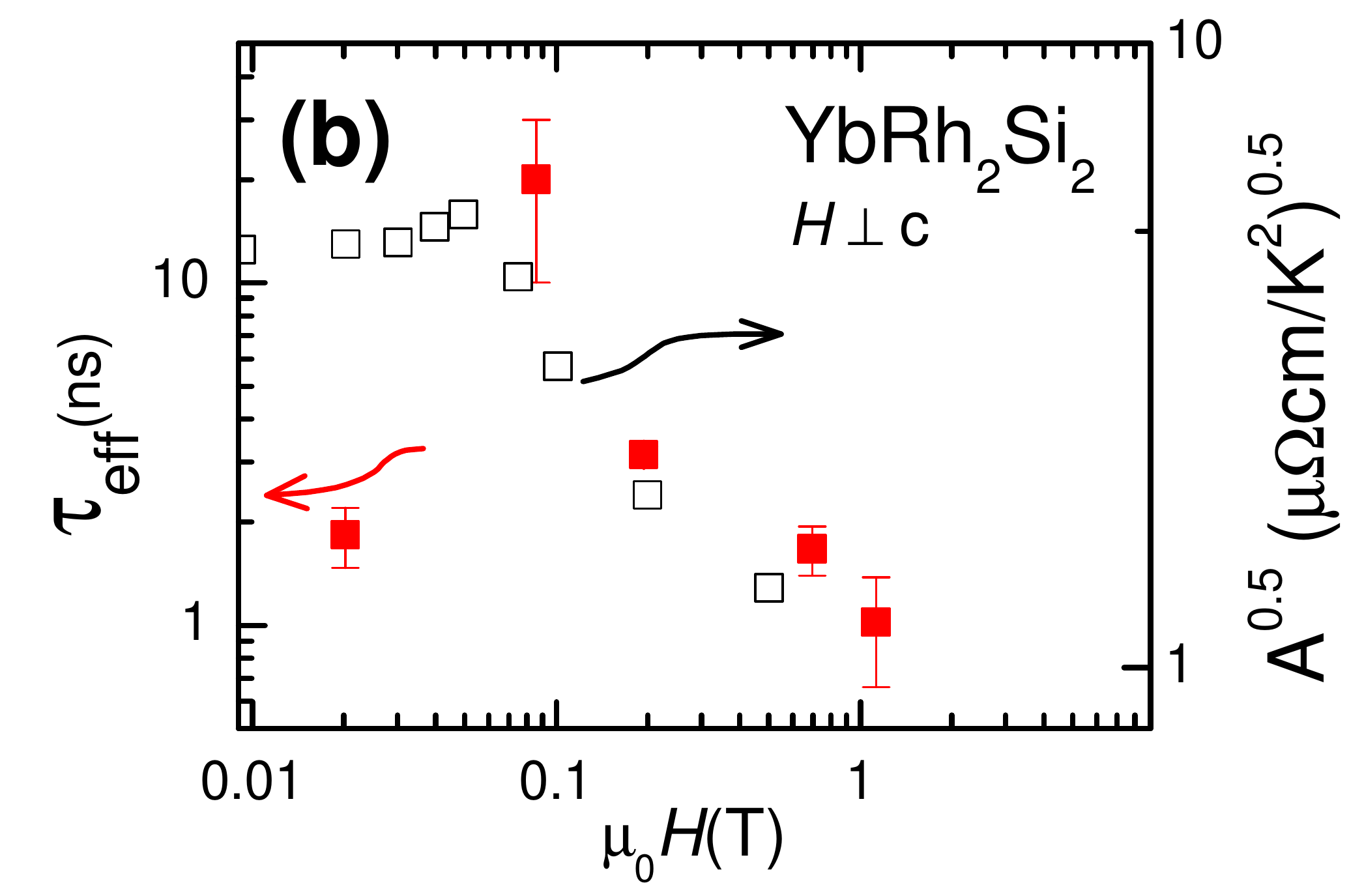}
%\end{center}%
\caption{%
Field dependence of (a) the ESR $g$ factor temperature variation, $\delta g=g(2\,{\rm K})-g(20\,\rm K)$ (data for $\mu_0H>1.8$~T from \cite{schaufuss09a}), and (b) the effective relaxation time $\tau_{\rm eff}(1.8{\rm K})=B_{\rm res}/(\nu\Delta B)$ (with $\tau_{\rm eff}(0.086\,{\rm T})$ obtained from linear extapolation of $\Delta B$ in \cite{duque09a}). Data are compared with the specific heat Sommerfeld coefficient $\gamma_0$ \cite{oeschler08a} and the quasiparticle scattering cross section $A$ \cite{gegenwart02a}, respectively.
}
\label{Fig3}
\end{figure*}
At all investigated fields and for $T\lesssim10$~K the linewidth shows a linear $T$ dependence, $\Delta B=a+b\cdot T$. The anomaly at $T^*$ (arrow) is found for the 34 GHz / 0.68 T data only. This is because the data at 0.02~T and 0.19~T are taken at $T>T^*$ and at 1.1~T the data accuracy is insufficient to resolve the anomaly at $T^*$. At low fields the slope $b$ varies slightly but not monotonous with field, see Fig.\ref{Fig2}c. Note in particular the slope of the 0.68~T data which is smaller than the slope of the 0.19~T data. The residual linewidth $a$ and, to a lesser extend, the slope $b$ roughly vary with $B^2$ as indicated by the dashed line. Although $a\propto B^2$ points towards an inhomogeneous broadening such behavior would also be consistent with a bottleneck relaxation of the coupled Yb$^{3+}$- conduction electron resonant collective mode as was identified from ESR investigations of Lu-doped YbRh$_{2}$Si$_{2}$ \cite{duque09a}.   
\section{Discussion}
The ESR of YbRh$_{2}$Si$_{2}$ clearly reflects the local magnetic properties of the Kondo ion Yb$^{3+}$ and should contain valuable information on the Kondo interaction between 4$f$ spins and conduction electrons.  
A suppression of the Kondo effect by magnetic field and temperature should therefore be visible in the ESR linewidth and in the $g$-factor. This, indeed, is qualitatively seen in Figs.\ref{Fig2}(a) and \ref{Fig2}(b) and was reported for high field values in Ref.\cite{schaufuss09a}. We quantitatively captured the $g$-factor behavior of Fig.\ref{Fig2}a by plotting the difference of the $g$ factor at low and high temperatures, $\delta g=g(2{\rm K})-g(20{\rm K})$. This is shown in Fig.\ref{Fig3}a and confirms the close relation to the Sommerfeld coefficient $\gamma_0$ which is a measure of $m^*$. Note the further increasing $\delta g$ at fields below the critical field $B_c=60$~mT, i.e. in a region with a long range, but weak antiferromagnetic ordering which has the signatures of a heavy LFL state according to $\rho(T,B)$ and $C(T,B)$ with $\gamma_0\simeq1.7$~JK$^{-2}$mol$^{-1}$.\\
Assuming that the observed resonance may be described by a heavy electron spin resonance a semiphenomenological Fermi-liquid description could explain the linewidth narrowing provided by hybridization, i.e. the linewidth $\Delta B$ should contain a factor $m/m^*$ (where $m$ is the free electron mass and $m^*$ is the effective quasiparticle mass) \cite{abrahams08a}. Hence, the effective spin relaxation time $\tau_{\rm eff}=B_{\rm res}/(\nu\Delta B)$ should be proportional $m^*$. Indeed, as shown in Fig.\ref{Fig3}b, $\tau_{\rm eff}$ at $T=1.8$~K increases with decreasing field, similar to the behavior of $\gamma_0\propto m^*$ and in agreement with the assertion of a Fermi liquid description \cite{abrahams08a}. However, this trend does not to seem to be continued for fields below the critical field $B_c=60$~mT. On the other hand such behavior is reminiscent to the magnetic field dependence of the quasiparticle scattering cross section $A$ that diverges at $B_c$ from both sides \cite{gegenwart02a} and that is proportional to  $m^{*2}$. As shown in Fig.\ref{Fig3}b a comparison of $\tau_{\rm eff}$ with $\sqrt{A}$ again suggests $\tau_{\rm eff}\propto m^*$. 
Such proportionality shows that the relaxation between Yb$^{3+}$ spins and heavy conduction electrons does not determine the linewidth because then an increasing $m^*$ would \emph{suppress} $\tau_{\rm eff}$ by a Korringa mechanism. This indicates a bottleneck relaxation of a strongly coupled Yb$^{3+}$- conduction electron resonant collective mode as supported by other experimental results \cite{wykhoff07a,duque09a}. The realisation of such a collective spin mode in the presence of a strong magnetic anisotropy and Kondo interactions will be described in a forthcoming paper.  
\section{Conclusion}
Our low-temperature ESR measurements at 34 GHz (0.68~T) of the Kondo lattice compound YbRh$_{2}$Si$_{2}$ confirmed a continuos decrease of the linewidth for temperatures upon cooling down to 0.6~K. Both $g$-factor and linewidth display a crossover behavior for the same temperatures where thermodynamic and static magnetic properties indicate a crossover to LFL behavior. Compared to ESR results at higher fields \cite{schaufuss09a} the crossover of the ESR parameters at 0.68~T appears more pronounced and in a much narrower temperature range. We confirmed the Kondo-like behavior of the field/temperature dependence of the ESR $g$ factor for fields down to 20~mT, i.e. smaller than the critical field $B_c=60$~mT. The field dependence of the low-temperature linewidth agrees with the behavior of the quasiparticle cross section.

\begin{acknowledgement}
We acknowledge support by the Volkswagen foundation (project I/82 203).
\end{acknowledgement}

% Use the following code if you wish to generate your bibliography with BibTeX;
% replace the string "pss-demo" below with the name(s) of
% the BibTeX data base(s) you want to use.
% The resulting bibliography-output (the contents of the .bbl file)
% must be pasted back into this file before submission.
%
%\bibliographystyle{pss}
%\bibliography{JoergBib}
\providecommand{\WileyBibTextsc}{}
\let\textsc\WileyBibTextsc
\providecommand{\othercit}{}
\providecommand{\jr}[1]{#1}
\providecommand{\etal}{~et~al.}

\end{document}